# A Comprehensive Study for Multi-Criteria Comparison of EV, ICEV, and HEV

Tushar Gahlaut*[a], Gourav Dwivedi[a]
[a]Indian Institute of Technology Delhi, India
Corresponding Author: Tushar Gahlaut*(Tushar.Gahlaut@dms.iitd.ac.in)

**Abstract**

Electric vehicles (EVs) are increasingly becoming popular as a viable means of transportation for the future. The use of EVs may help to provide better climatic conditions in urban areas with a pocket friendly cost for transportation to the consumers throughout its life. EVs enact as a boon to the society by providing zero tailpipe emissions, better comfort, low lifecycle cost and higher connectivity. The article aims to provide scientific information through the literature across various aspects of EVs in their lifetime and thus, assist the scholarly community and various organisations to understand the impact of EVs. In this study we have gathered information from the articles published in SCOPUS database and through grey literature with the focus of information post 2009. After identification of various factors while purchasing the vehicle, the total cost of ownership (TCO) for vehicles is calculated and their average TCO for each segment is considered for the study. Following that, we investigate the ranking of EVs, vehicles powered by internal combustion engines (ICEVs), and hybrid electric vehicles (HEVs) in a variety of price segments by employing a combination of two different multi-criteria decision making (MCDM) techniques. Initially, best-worst method (BWM) is used to determine the weights of each of the identified criterion, which is thereafter, used in conjunction with the technique for order preference by similarity to ideal solution (TOPSIS) to compute the rank of the available alternatives using BWM. The ranking obtained clearly indicates that EVs should be first purchase choice of the consumers, followed by HEVs and ICEVs respectively. Thus, the results help us conclude that EVs enact as a sustainable means of transport for the future.

**Acknowledgements:** This article would not have been possible without the appropriate support and input from the academic and industry experts. We are also grateful to our peers for their insightful comments and continuous guidance by Prof. B.K. Panigrahi, that helped in improving the study.

1. **Introduction**

With the growth of urbanisation and globalisation there has been a rapid increase in transportation sector over the years. This has led to the rise of transportation industry across all segments. Though different modes of transportation have made the mobility of goods and individuals easier, but simultaneously increased the increased global green-house gas (GHG) emissions and burden on the pockets of individuals (Geurtsen & Wilford, 2009). The emissions through vehicles have made typical climatic conditions to live across the urban areas around the globe. India ranks third after China and USA in emitting GHG emissions and accounts for 2.9 $GtCO_2e$ per year (Gupta R. et al., 2022, p. 36). The transportation sector alone accounts for 9% of these GHG emissions (Gupta R. et al., 2022, p. 36). The time has come to adopt eco-friendly technologies for mobility. Innovative people are keen to adopt these technologies at early stages. As a result, the governments across the globe are taking steps to enhance the adoption of EVs and HEVs. The governments are simultaneously enhancing the use of ethanol blended petrol (EBP) as fuel for all the gasoline-based vehicles. Nowadays, gasoline typically contains E10 (i.e. mixture of gasoline and ethanol containing 90% gasoline and 10% ethanol).
.

The currently available categories of vehicles based on fuel-type constitutes ICEVs (gasoline-based vehicles), EVs and HEVs (i.e. mild hybrid and strong hybrid). From these available alternatives consumers' view EVsas both cost-effective and environmentally beneficial.. The growth of EVs can be attributed to various factors, including advancements in battery technology, government incentives, and a growing awareness of environmental sustainability. As significant automakers invest heavily in research and development, the market for EVs has expanded, offering consumers a diverse range of electric options. Increasing concerns about air quality, climate change, and the desire for energy independence have further accelerated the transition towards EVs. Thus, they are expected to reduce the GHG emissions up to a great extent and thus, provide better quality of air to breathe in for humans as well as other living beings. EVs are quite and does not have any fuel cost associated with them for their operation (Un-Noor et al., 2017) contradicting their counterpart of internal combustion engine vehicles (ICEV). IF 1,50,000 km is assumed as the average lifespan of EVs it has been found that current electricity mix can reduce the GHG emissions by 10-24% (Hawkins et al., 2013). If the electricity is generated from carbon free sources it has been found that electric vehicles will be responsible for 6.85% of total GHG emissions (which is currently 55.2% from conventional cars) and 5.76% of total air emissions (which is currently 61.4% from conventional cars) (Nanaki & Koroneos, 2013). Also, it has been found that 59-62% of consumed energy by EVs can be converted to power which is far more than their conventional counterpart that can convert only 17-21% of consumed energy to power (Wei & Dou, 2023). According to Nitin Gadkari, Minister of Road Transport and Highways (MoRTH) the government of India (GOI) has set an ambitious target to have EV sales penetration of 30% for private cars, 80% for 2-wheerers and 3-wheelers and 70% for commercial vehicles by 2030. These facts can enact as a motivating factor amongst the consumer to shift from ICEVs to EVs. EVs are considered as eco-friendly mode of transport which are convenient, self-reliant and pocket friendly. Thus, it can be illustrated that the adoption of EVs is boon to the society. Environmental awareness among the society members enacts as a major factor to enhance EV adoption throughout the world (Nguyen & Pojani, 2023). This article lays emphasis on the pros of adopting EV technology and thus, prohibiting the use of traditional ICEVs.

The higher purchase cost of EVs and the misconception among the individuals for their shorter range, charging inconvenience and lack of safety inhibits the individuals for not switching to EVs. Thus, the government takes the liability to enhance EV adoption by framing various policies in their country. The government across various nations are making numerous policies including green licence plates for both private and commercial vehicles to exempt them from certain permit rules, subsidies and tax rebates to boost the adoption of EVs in their country. These policies may vary across states in a country. These policies can enact as an enabling factor to spread awareness among people for the advantages of EVs and thus, encourage EV adoption within the country.

In near future it is expected that newer business models will come into play and EVs can conveniently enact as an efficient energy storage device for efficiently implementing the newer business models. This stored energy can be used for developing vehicle-to-home (V2H) or smart homes and thus, powering the equipments in the household if required (Weiller & Neely, 2014). Thus, it can be illustrated that EVs will not only be used for transportation and logistics devices rather enact as energy service providers as well. This is an added advantage of EVs over the conventional gasoline vehicles since they can never be used as energy service providers.

Based on the objective of our study we have identified various research questions (RQ). The RQ's are as follows-

**RQ1:** What are the factors that should be considered by the buyers while purchasing a vehicle?

**RQ2:** What should be the desirable weights of each of the identified factors/criterion's to make appropriate decision of vehicle purchase?

**RQ3:** What should be the preferred choice of the vehicles from the available alternatives of EVs, ICEVs and HEVs?

The remaining article is organised as follows: section 2 indicates the impact of electric vehicles, section 3 indicates total cost of ownership (TCO) of EVs, section 4 depicts EVs are more convenient, section 5 describes the self-reliance of EVs, section 6 describes the factors for comparison among EV, ICEV and HEV, section 7 describes the methodology while section 8 comprises the results and section 9 conclusion.

2. **Impact of Electric vehicles**

*2.1 Environmental Impact*

The life cannot exist without transportation especially in the urban regions, but the conventional combustion engine is rapidly becoming antiquated. Vehicles powered by gasoline or diesel emit a great deal of pollution, and fully electric vehicles are rapidly taking their place. Fully electric cars (EVs) are significantly better for the environment and have no exhaust emissions. Compared to conventional vehicles, electric and hybrid vehicles can offer significant pollution advantages. When used exclusively in electric mode, plug-in hybrid electric vehicles (PHEVs) emit no tailpipe emissions at all. Benefits of HEV emissions differ depending on the type of hybrid power system and vehicle model (U.S. Department of Energy). Electricity is used by electric vehicles to charge their batteries rather than fossil fuels like gasoline or diesel. Since electric vehicles take less energy to charge than gasoline or diesel to meet your travel needs, charging an electric vehicle is more cost-effective than filling it up (Niti Aayog). The electric car is an alternative to satisfy the demand for a green source of transportation with lower emissions and improved fuel economy. It helps to mitigate the effects of rising fuel prices and to adopt environmental legislation with higher requirements. The way that the nation generates its electricity has a direct bearing on the environmental effects of EVs. EVs may not be useful in reducing greenhouse gas (GHG) emissions in nations lacking an environmentally favourable mix of electricity generation (Woo et al., 2017). A major factor in assessing the environmental effects of EVs is the reduction of $CO_2$ gas emissions. When compared to ICE vehicles, the extremely efficient electric motors in EVs contribute to their lower CO2 gas emissions. Driving an electric car can be more environmentally friendly if renewable energy is used. If renewable energy sources, like solar panels, are placed at home for charging, the cost of electricity can be further decreased. EVs also make a great deal less noise, which helps significantly lessen sound pollution, especially in cities. Electric vehicles have the silent functioning capability as there is no engine under the hood. Thus, they emit zero noise. The electric motor functions so silently that you need to peek into your instrument panel to check if it is ON. Electric vehicles are so silent that manufacturers have to add false sounds in order to make them safe for pedestrians (Niti Aayog). EVs have the potential to drastically reduce greenhouse gas emissions from the transportation sector, which presents opportunities and huge environmental benefits. The fact that EVs help to cut greenhouse gas (GHG) emissions is one of the key reasons for their rising popularity. When fuels are burned directly in internal combustion engines (ICEs), hazardous gasses including carbon dioxide and carbon monoxide are released. Despite having internal combustion engines, HEVs and PHEVs produce fewer pollution than conventional cars. However, other ideas contend that the electricity used by EVs may increase greenhouse gas emissions from power plants, which must generate more energy due to the additional load that EVs provide. This argument can be supported by the observation that peak load power plants are most often of the ICE type, or that they can generate power using coal or gas. EVs will cause these facilities to operate and produce CO2 emissions if they generate excess load during peak hours (Ma et al., 2012).

According to Sioshansi et al. (2011), EV adoption will result in a greater CO2 production than ICEs when power generation from coal and natural gas is used. But not all of the power comes from these sources. Many alternative power-generating technologies exist that emit fewer greenhouse gases. When they are taken into account, the amount of greenhouse gas produced by power plants as a result of EV penetration is lower than that produced by ICE vehicle-based equivalent power generation. Additionally, the power plants generate energy in large quantities, reducing emissions per unit. Emissions from the transportation sector and power generation can be decreased with the appropriate integration of renewable sources, which EVs can greatly support (Kutt et al., 2013). EVs emit fewer emissions throughout the course of their lifetime than do traditional cars. EVs have a lower value for this measure, which is also known as well-to-wheel emission (Donateo et al., 2014). Denmark used EVs and electric power to reduce transportation-related CO2 emissions by 85%.

EVs have a big potential to affect the electrical grid, the environment, and other connected areas. If EV penetration increases to a certain point, the current power grid may experience severe instability. However, with careful management and cooperation, EVs can be transformed into a significant factor in the smart grid concept's effective execution (Un-Noor et al., 2017). Given the state of the transportation industry today, it is anticipated that EVs will soon have a significant market share due to their rapid development rate in distribution networks. Because of the extensive charging usage of EVs, which is negatively impacting the current conventional distribution grids, the current power networks may experience extra loads.

The widespread installation of EVs in urban areas can lead to the electrification of the transportation industry. Because of the transportation sector's development, there is a more hospitable atmosphere thanks to lower CO2 emissions. According to Shuukat et al. (2018), there will be a reduction in carbon dioxide emissions of up to 1-6% until 2025 and 3-28% until the end of 2030. By integrating EVs into power networks, this is accomplished. However, widespread use of EVs with cutting-edge technologies have positive effects for the environment and the economy. Furthermore, a large-scale EV integration into the V2G (vehicle to grid) environment advances the clean and safe energy society. Improvements in electric vehicle (EV) technology can lessen reliance on fossil fuels and create a greener environment (Habib et al., 2018). In a clean energy environment, V2G technology is essential ((Habib et al., 2018)).

*2.2 Sustainability*

In addition to reducing greenhouse gas emissions, plugging in electric vehicles (EVs) can provide many amazing features like load balancing, reactive power support, active power regulation, and sustainability for renewable energy resources (Shaukat et al., 2018; Ashique et al., 2017).

Although the sophisticated batteries found in electric cars are made to last a long time, they will inevitably degrade. Eight-year/100,000-mile battery guarantees are being offered by a number of electric car manufacturers. The National Renewable Energy Laboratory of USA using its predictive modelling suggests that modern batteries could endure 12 to 15 years in temperate regions and 8 to 12 years in harsh ones (U.S. Department of Energy). The vehicle-battery-environment thermal system, driving and charging habits, battery cell chemistry, and design are additional elements that affect battery life in addition to climate. While replacement battery prices have not been disclosed by manufacturers, some are charging monthly fees for extended warranty plans. This is because when the batteries require replacement after the guarantee has expired, the cost could be substantial. As battery technology advance and production volumes rise, it is anticipated that battery prices will continue to decline. The complete recycling of lithium-ion batteries is a challenge, as only a small number of firms are able to accomplish this. Lithium-ion cells, unlike the earlier nickel-metal and lead-acid ones, are not composed of caustic chemicals, and their repurposing can lessen the demand for "peak lithium" or "peak oil" (Shareef et al., 2016). A study by Mckinsey indicates that EV battery cell production from recycled raw material reduces $CO_2$ emissions by 28% as compared to cell manufacturing from virgin material (Breiter A. et al., 2023, p. 4). Till the EV technology is about to mature the EVs will become a sustainable means of transport and enhance the efficiency and transparency in the supply chain.

### 3. TCO for Various Types of Vehicles

A viable replacement for the present generation of fossil fuel-powered automobiles are electric vehicles (EVs). However, just 4.6% of EV sales worldwide were made as of 2020 (Huang et al., 2019). The percentage of EV sales in global auto sales has climbed to 14% in 2022 (International Energy Agency, 2023). Due to the great efficiency of electric-drive components, fuel costs can be significantly reduced by electric vehicles. The operational cost of EVs have also been reduced across various nations due to various subsidies on charging for electric vehicles (Palmer et al., 2018). Since PHEVs and all-electric cars depend entirely or partially on electric power, their fuel efficiency is calculated differently from that of traditional cars. Common measures include miles per gallon of gasoline equivalent (MPGe) and kilowatt-hours (kWh) per 100 miles (U.S. Department for Energy). Modern light-duty all-electric cars (or plug-in hybrid electric vehicles, or PHEVs) can achieve over 130 MPGe and go 100 miles on 25–40 kWh depending on how they are operated. Also, EVs have very low maintenance costs because they don't have as many moving parts as an internal combustion vehicle. The servicing requirements for electric vehicles are lesser than the conventional petrol or diesel vehicles. Therefore, the yearly cost of running an electric vehicle is significantly low (Niti Aayog). A study indicates that the daily operational cost of conventional vehicles is 2.5 times than that of electric vehicles. Compared to conventional cars, HEVs and EVs provide a lower carbon and environmental footprint; yet, their fleet share in most vehicle markets is currently too small to have a meaningful impact (Palmer et al., 2018). Until 2025, ICEVs are probably the most cost-effective technology in some TCO scenarios, such as the short-distance ones; however, in other scenarios, like the long-distance ones, EVs may still be more cost-effective than ICEVs by that year (Wu et al., 2015). This is mostly because EVs have lower operational costs per kilometer than do conventional cars. However, in every scenario, the capital cost of an electric vehicle (EV) is more than that of a conventional car. An extended driving range raises the operating cost's weight in the TCO computation, which boosts EVs' relative cost efficiency. Smaller cars with comparatively lower capital costs also consistently result in a greater operating cost weighting in the TCO computation, which raises the comparative cost efficiency of EVs (Wu et al., 2015). A study by Mckinsey indicates that TCO for EV two-wheelers is much lower than their traditional counterparts while, the TCO for compact SUV in the EV segment will be lower than their ICEVs counterparts (Gupta R. et al., 2022, p. 73).

It has been found that compact EVs require only 60% of fuel while light trucks require 64-78% of the fuel (Harvey, 2020). The future cost of operating EVs is expected to further reduce up to a great extent due to advancement in EV technology and higher use of renewable sources for electricity generation at lower costs (Harvey, 2020; Murphy et. al, 2019). The electricity cost by the renewable sources of energy is decreasing continuously and is expected to decrease further upto a great extent (Murphy et. al, 2019). According to the analysis by Al-Alawi & Bradley, (2013) variables including incremental cost, gas prices, and annual driving distance—all of which have been thoroughly studied in the literature—have an impact on both TCO and payback period. For instance, a 20% increase in gas costs is demonstrated to result in a 31% reduction in the mid-sized PHEV20's payback period when compared to a CV. It was also demonstrated by this analysis that salvage value, maintenance expenses, and fuel economy—three relatively understudied aspects of TCO modelling—have an impact on TCO and payback period.

The cost of the battery pack, which is substantially more expensive than the drivetrain of an ICEV, is the main factor for the high purchase price of BEVs. Future battery prices are predicted to drop at a rapid rate, which will increase the competitiveness of BEVs in terms of TCO and purchase price. BEVs' cheap ongoing ownership costs could have an intriguing impact on their resale value and longer ownership durations (Hangman et. al., 2016). Comparable used ICEVs and HEVs may have substantially reduced TCO when a used BEV is purchased instead of a new one because of the absence of the high rate of depreciation and financing expenses. These are elements that may have a favourable impact on BEVs' resale value, which would also make new BEVs more desirable. Since operating costs account for a growing portion of the total cost of older cars, a similar impact could potentially make BEVs more competitively priced over longer ownership periods (Hangman et. al., 2016). However, the short battery life and replacement cost may negatively impact BEVs' long-term TCO and resale value.

The Total Ownership Cost (TOC) of an EV encompasses various elements beyond the initial purchase price, including maintenance, charging infrastructure, electricity costs, and potential incentives. Accurately estimating these costs is paramount for individuals and businesses considering the switch to EVs, as it provides insights into electric transportation's long-term affordability and economic viability.

Considering the case of various states across different nations it has been found that buying an electric vehicle entails lower registration costs and road tax than buying a gasoline or diesel car. In order to reduce the purchase burden of vehicles on the consumers the governments across various nations have come up with numerous subsidy programs for the consumers (Harvey, 2020). This step by the government is pushing the consumers to enjoy low fuel economy vehicles with higher efficiency and simultaneously enabling them to contribute for the greener environment. The policies and incentives offered by the government varies across states in a country like India. In this study, we have estimated the TCO for a vehicle owner in India for top-selling cars in EV, ICEV and HEV segment. Thereafter, average TCO for various vehicle models in different price segment for EVs, ICEVs and HEVs is used to compute our results. The detailed results of TCO along with the various assumptions undertaken for calculation of TCO are depicted in the *Appendix.* -

4. **Convenience**

EVs are considered to be more comfortable due to lack of transmission system. The lack of transmission system and the availability of electric motor with high torque makes the driver enjoy with dynamic and powerful acceleration. This makes it easier for the driver to pass or merge with the highway ("pleasure of driving an electric vehicle - Easy electric life - Renault group," 2020). Also, EVs are easy and convenient to drive and control since they don't have gears. The user only has to sit and accelerate without making a combination of clutch and acceleration. There are no complicated controls, just accelerate, brake, and steer.

Since electric vehicles are equipped with an electric motor and a battery at the bottom of the surface also, the non-availability of internal combustion engine provides a lower centre of gravity thus, providing improved handling, stable ride and higher responsiveness (Union of Concerned Scientists, 2018; Admin, 2023). EVs are provided with regenerative braking system that enables the EVs to convert the stored mechanical energy into electrical energy and transfer it to the battery pack. Thus, provide the EVs with the ability to recharge themselves up-to a certain extent and so increasing the efficiency of the vehicles (Admin, 2023). This makes driving smoother.

Due to the lack of internal combustion engine and the availability of contactless transmission system the EVs make zero noise thus, providing peace and making them appealing to our ears (Geurtsen & Wilford, 2009). Inside the cabin one can easily enjoy the peace and listen to the radio or passenger chats more conveniently ("pleasure of driving an electric vehicle - Easy electric life - Renault group," 2020). As mentioned earlier, zero noise by EVs also lead to reduction in noise pollution levels.

Also, in EVs the user may not have to rush to a refuelling station as in case of gasoline-based vehicles. This is because the user may simply charge the vehicle at their home by just plugging it in a few seconds and lets the user wake up in the morning with a "full tank" (Union of Concerned Scientists, 2018). Consider yourself at a crowded gas station at rush hour, running out of time to get to work. An electric car is a simple solution to these issues. Just leave your car plugged in for four to five hours at home before you intend to leave. It is quite convenient to arrange your trips ahead of time if you can find a charger where you park at home (Niti Aayog). What happens if you ever forget to plug in your computer? Then, if you are riding a two-wheeler on the road, you may readily enlist the aid of rapid chargers or even battery changing services. When you want to charge your vehicle, just plug it in to a home or public charger. Efforts are being taken up by government across various nations of building charging spots at various work facilities and parking lots. Thus, making it more convenient for the user to charge their vehicle and avoid running to refuelling stations to recharge their vehicles. This, charging system is quite convenient and hassle free for the users using vehicles for local commute (which is a common practice).

Due to the availability of fewer moving parts the EVs experience lower wear and tear of various parts. This provides a longer life for EVs and thus, providing them with lower maintenance needs. This makes it convenient for the user to spend less money and time to maintain their vehicles.

5. Safety

The advanced automotive electronics technology systems are helping to build an intelligent system for transport (Shi et al., 2022). In the transportation sector, electric vehicles (EVs) are starting to use this smart system initially. This has provided the vehicles with advanced safety and security systems as compared to conventional vehicles. The EVs have an added advantage over ICEVs since they are provided with blind spot monitoring system, collision avoidance system, cybersecurity systems, assisted lane change system and automotive regenerative braking system (Admin_noodoe, 2023).

There is a misconception that EVs are less safer than their counterpart ICEVs and HEV in the same price segment. The lack of engine in EVs provide them with higher crumple zones to absorb the impact of collisions. Blind-spot monitoring system improves the visibility and intimates the driver about the vehicle in the blind spot thus, avoiding collisions. Collision avoidance system enables the automotive regenerative braking system to automatically reduce the speed of the vehicle and apply breaks if needed to avoid collision with other vehicles. This help to reduce the road accidents and thus, the injuries. EVs are less prone to accidental fires. This is because EVs don't run on flammable material and also, the battery material is provided with battery casing as well as vehicle casing thus, providing double insulation layer for temperature rise in case of a crash. Also, the battery management system (BMS) in EVs and continuously monitors the battery performance help to control voltage and thermal levels of the battery (Hossain Lipu et al., 2021). The assisted lane change system in EVs enables the driver to make more informed decision and warn him/her to change/not to change the lane thus, avoiding collisions and thus, enjoy better and smooth driving experience.

## 6. Factors for Comparison of EV,ICEV and HEV

In this dynamic environment one finds it very difficult to decide on how to choose between an EV, ICEV and HEV. To make it convenient for the upcoming users to select EV or ICEV we have come up with various factors after discussing with the current EV users and certain policy makers. The various factors that one may consider while choosing among EV, ICEV and HEV are:-

- **Cost of ownership:** It is one of the primary factors considered by the buyer while purchasing a vehicle. The buyer tends to have the lowest cost of ownership for the vehicle he/she tends to purchase while simultaneously taking the other factors into consideration.

- **Re-fuelling infrastructure and convenience:** All the vehicle buyers give a thought towards the distance they may drive regularly in a day. Once the user is aware of the average commute distance on daily basis he/she looks after the availability of refuelling infrastructure for their vehicle. If the person is satisfied with the available refuelling infrastructure to fulfil his/her refuelling needs then he/she may choose to opt the vehicle. The individual may also try to give thought towards the re-fuelling convenience of the vehicle. EVs are convenient to re-fuel for the users since they can be charged at home while ICEVs and HEV are to be taken to a re-fuelling station for re-fuelling.

- **Range:** The average distance the vehicle covers when its fuel capacity is completely filled is determined as the range of the vehicle. The user tends to have maximum range for the vehicle at minimum cost in such a manner that his daily commuting needs can be easily covered with the available refuelling infrastructure.

- **Safety & Comfort:** Safety is one of the minor concern for most vehicle buyers. The buyers tend to invest more while considering the safety aspect for the vehicle. Comfort is one of the secondary factors that one may consider while purchasing a vehicle. The buyer tends to prefer the vehicle with highest level of comfort in a particular price band but simultaneously considers other factors as well as per his/her needs.

- **Network Effect:** It is the general tendency of a human to purchase the vehicle from the vehicle segment (EV, ICEV and HEV) which is likely to sustain in the market for one reason or the other. The buyers develop the confidence for the segment to be sustainable if they find the market confidence with a proper network of vehicle models in the desired price segment (Miao et al., 2017).

- **Environmental impact:** The buyers rarely consider this factor while purchasing the vehicle while the present government officials develop their policies considering this as one of the key factors for EV/ICEV/HEV adoption. This may be considered by the individuals to develop a healthy living environment. The environmental impact factors generally include the amount of emissions, the efficiency of energy consumption by the vehicle etc.

- **Policy Push and Regulations:** One tend to look after the guidelines/policies provided by the government to promote/demote a particular vehicle segment (i.e. EV or ICEV or HEV). The buyer usually considers the policy push by the government as one of the most important factors while purchasing the EV. The policies and regulations motivate the buyer to buy the vehicle from a particular segment without compromising on his/her needs.

## 7. Methodology

The previous research on electric vehicles majorly focusses towards their environmental and techno-economic feasibilities while very little is discussed about the advantages and disadvantages of electric vehicles, which indicates that reviews are mandatory this theme. We have used the articles from SCOPUS database rather than Web of Science and Google Scholar since it is a more comprehensive database and covers information from a wider range of sources. We have used the combination of relevant keywords using "AND", "OR" and "NOT" operators in the database in the title and abstract. We have used the keywords "electric vehicl*", "advantag*", "disadvantage*", "merits" and "demerits". The word's asterisk at the end denotes the inclusion of all occurrences of that word; for eg., vehicl* comprise of both vehicle and vehicles. A total of 869 relevant articles were published from 2009 to 2023. There was an annual increase in the number of publications every year. We only considered the research and review articles from journals and conference proceedings (which were 424 in the database). The articles were filtered out as per relevance and information. This research did not include articles from conference proceedings, short surveys, letter, books and book chapters. The process for selection of articles has been described in Figure 1. We also explored certain information from grey literature which includes government sources, reports and information available on company websites was also included.

Also, we used hybrid MCDM (i.e. BWM-TOPSIS) model to identify the best alternative among EVs, ICEVs and HEV (as described in Figure 2.) in the price range varying from 8.00 lakhs to 25.00 lakhs. In order to avoid confusion, we have considered only petrol-based vehicles under the ICEV segment and strong hybrids under the HEV segment. In this study, we have segmented various EVs', ICEVs' and HEVs in different price segments. We consider various factors/criterion which the user may consider while purchasing a new vehicle. These factors are obtained after discussion with various experts and users. After identifying the decision-factors we interact with the users to rate each of these criterion w.r.t. the best criterion and worst criterion (best and worst criterion's are selected by the users). Thereafter, we apply best-worst method (BWM) to obtain weights of each criterion. We gathered this information from different users and thereafter, took the average of weights that is used further to applied to finalise the best alternative using TOPSIS. The results and the literature enabled us to understand both the advantages and disadvantages of EVs in detail.

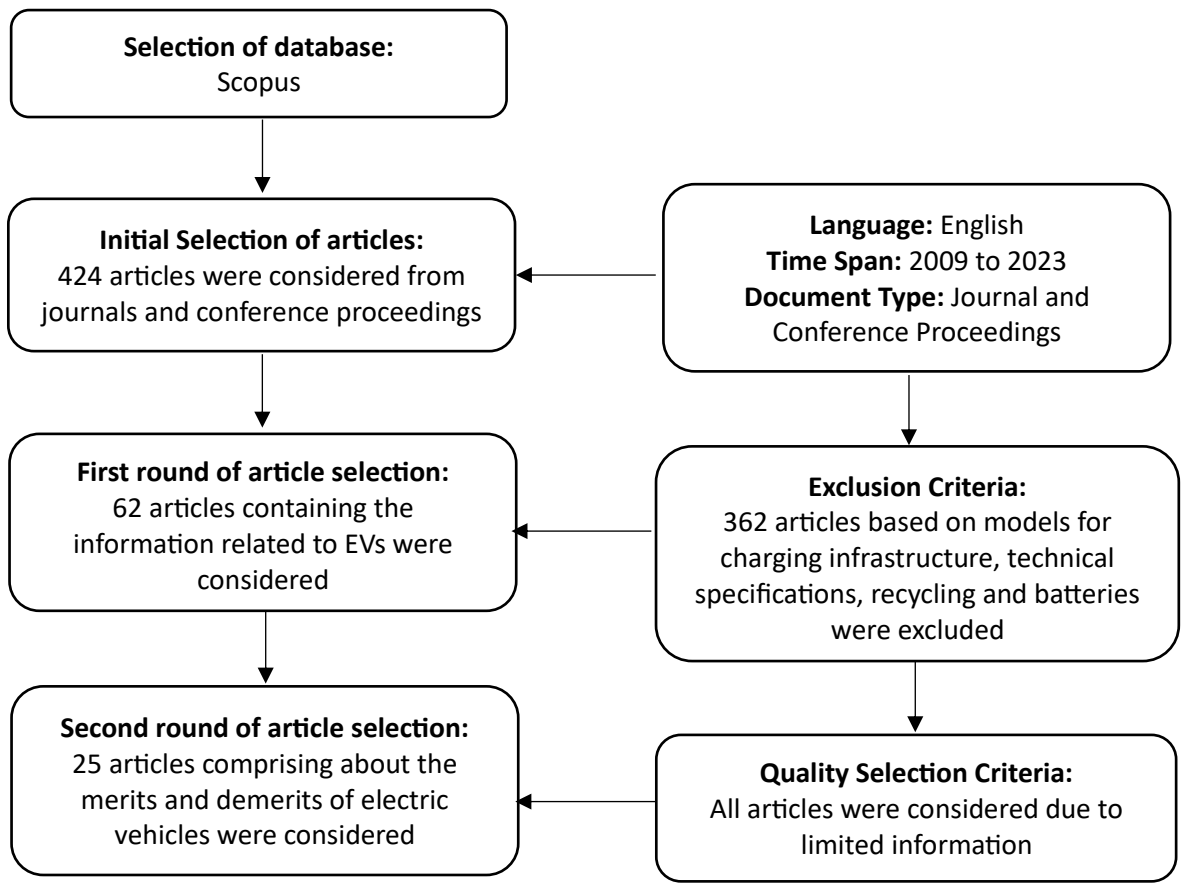

Figure 1.

*7.1 Best-Worst Method Procedure*

In 2015, Rezaei created a novel MCDM approach known as the best-worst method (Rezaei, 2015). This method is likewise based on pairwise comparisons of the criteria and sub-criteria, just like AHP (Analytical Hierarchy Processing). In comparison with AHP, this method is thought to be more effective. Comparisons in AHP are expressed as "n(n-1)/2," whereas in BWM they are expressed as "2n-3." The computational steps for BWM are as mentioned below:-

1. **Step 1:** Establish the group of different criteria ($C_1$, $C_2$, $C_3$, ------- ,$C_n$).

Where 'n' is the number of criteria

2. **Step 2:** Determine which criteria is the best (more significant) and which is the worst (less significant) by consulting experts. Select one at random if there are several criteria that are thought to be best or worst.

3. **Step 3:** Using numbers 1 through 9, ascertain which criteria is preferred over the worst and which is preferred over the other criteria. OW(others-to-worst) = ($a_{1W}$, $a_{2W}$, $a_{3W}$, --------, $a_{nW}$) and BO(best-to-others) = ($a_{B1}$, $a_{B2}$, $a_{B3}$, --------, $a_{Bn}$). are the vectors, respectively.

4. **Step 4:** Find the optimal weights of each criterion. The optimal weight vector is $W^* =$ ($W_1^*$, $W_2^*$, $W_3^*$, -------, $W_n^*$). The weights obtained can be considered as optimal weights if and only if the following conditions are satisfied: 1) $W_B/W_j - a_{Bj} = 0$ ; and 2) $W_j/W_W - a_{jW} = 0$. Thus, to satisfy these conditions minimization of the maximum absolute differences of $| W_B/W_j - a_{Bj} |$ and $| W_j/W_W - a_{jW} |$ is performed for all j. The optimization model is as follows :

**Model 1**

$$\text{Minmax}_j \left\{ \left| \frac{W_B}{W_j} - a_{Bj} \right|, \left| \frac{W_j}{W_W} - a_{jW} \right| \right\}$$

Subject to:

$$\sum_j W_j = 1 \quad \ldots\ldots (1)$$

$$W_j \geq 0, \quad (\forall j \in [1,n]) \quad \ldots\ldots (2)$$

This model can be transformed to Model 2.

**Model 2**

Min.  $\xi$

Subject to:

$$\left| \frac{W_B}{W_j} - a_{Bj} \right| \leq \xi \quad (\forall j \in [1,n]) \quad \ldots\ldots (3)$$

$$\left| \frac{W_j}{W_W} - a_{jW} \right| \leq \xi \quad (\forall j \in [1,n]) \quad \ldots\ldots (4)$$

$$\sum_j W_j = 1 \quad \ldots\ldots (5)$$

$$W_j \geq 0 \qquad (\forall j \in [1, n]) \qquad \ldots\ldots (6)$$

Thus, the minimum absolute difference $\xi^*$ and the optimal weights ($W_1^*$, $W_2^*$, $W_3^*$, ------ , $W_n^*$) can be obtained by solving model.

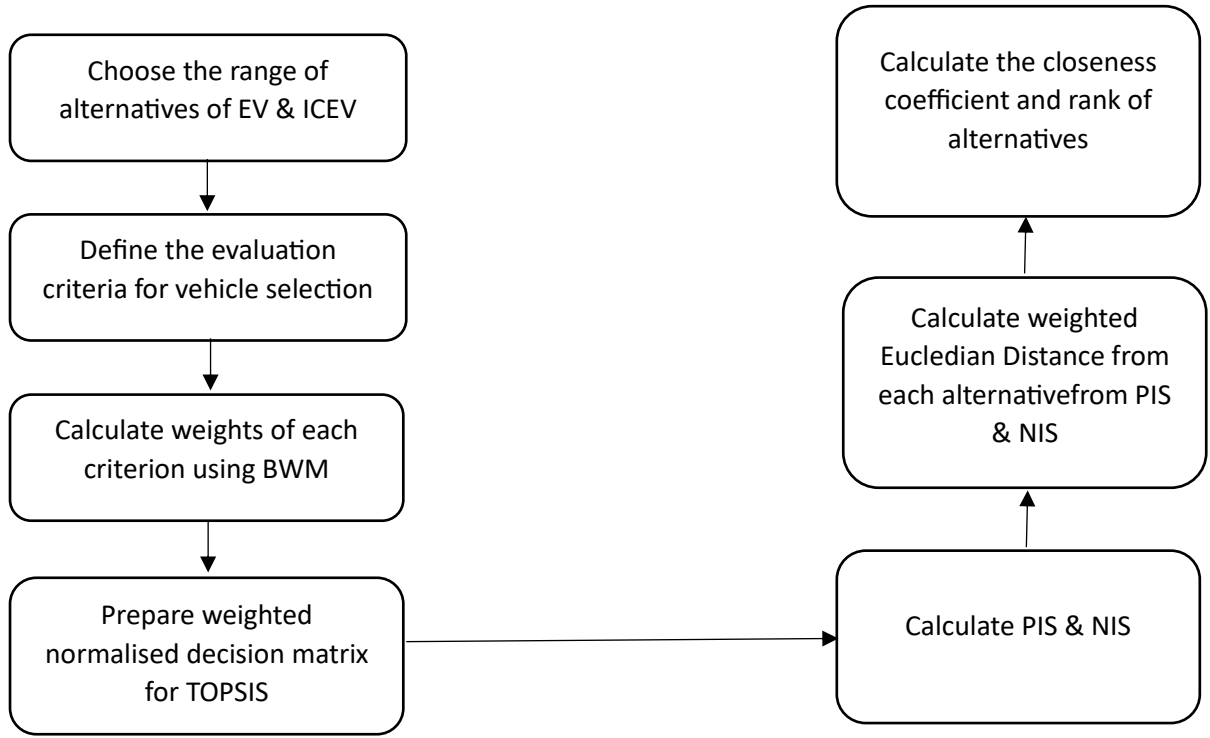

Figure 2.

*7.2 TOPSIS Procedure*

In theory and practice, TOPSIS is a straightforward ranking system that was created by Hwang and Yoon in 1981. Using the traditional TOPSIS method, one seeks solutions that are as far from the ideal negative solution as possible but also as close to the ideal positive response as feasible. In addition to offering a cardinal ranking of options and utilizing attribute information to the fullest, TOPSIS does not require individual attribute preferences (Chen and Hwang, 1992; Yoon & Hwang, 1995). In order to utilize this method, attribute values need to be numerical, exhibit a monotonic increase or decrease, and possess equivalent units. The computational steps for TOPSIS are as mentioned below:-

1. **Step 1:** Establish the group of different alternatives.
2. **Step 2:** Create normalised decision matrix

$$y_{ij} = \frac{x_{ij}}{\sqrt{\Sigma x_{ij}^2}} \quad \text{for } i = 1,\ldots,m \text{ and } j = 1,\ldots,n \qquad \ldots\ldots (7)$$

3. **Step 3:** Create weighted normalised decision matrix.

$$v_{ij} = w_j y_{ij} \qquad \ldots\ldots (8)$$

where $w_j$ is the weight of j criterion

4. **Step 4:** Calculate PIS (positive ideal solution; $V_{j+}$) and NIS (negative ideal solution; $V_{j-}$).

$$V_{j+} = \{v_1^*, \ldots, v_n^*\}, \ldots.. \text{ positive ideal solution}$$

Where $v_i^* = \{\max(V_{ij}) \text{ if } j \in J; \min(V_{ij}) \text{ if } j \in J'\}$

$$V_{j-} \text{ OR } A' = \{v_1', \ldots, v_n'\}, \ldots.. \text{ negative ideal solution}$$

Where $v_i' = \{\min(V_{ij}) \text{ if } j \in J; \max(V_{ij}) \text{ if } j \in J'\}$

5. **Step 5:** Determine the separation measure for each alternative

    Separation Measure for Positive Ideal Solution (PIS):

$$S_i^* = \left[\Sigma(v_i^* - v_{ij})^2\right]^{\frac{1}{2}} \quad i=1,\ldots,m$$

    Separation Measure for Negative Ideal Solution (NIS):

$$S_i' = \left[\Sigma(v_j' - v_{ij})^2\right]^{\frac{1}{2}} \quad i=1,\ldots,m$$

6. **Step 6:** Calculate the performance measure

$$C_i^* = S_i' / (S_i^* + S_i'), \qquad 0 < C_i^* < 1$$

Select the alternative with the value of $C_i^*$ closest to 1.

7. **Step 7:** Determine the rank of each alternative as per the performance measure.

## 8. Results

In this study we have used BWM to calculate the weights of each criterion considered to select the best alternative from the available range of vehicles. We took the average weights obtained from 15 datasets which is further used to evaluate the best available alternative of EV, ICEV and HEV (using both BWM and TOPSIS) in a particular price bracket.

| | |
|---|---|
| **Cost of Ownership** | 31.65 |
| **Safety & Comfort** | 5.45 |
| **Range** | 10.46 |
| **Network Effect** | 10.60 |
| **Refueling Infrastructure & Convenience** | 10.15 |
| **Environmental Impact** | 3.87 |
| **Policy Push & Regulations** | 27.82 |

Table 1. Criterion Weights using BWM

The weights for each criterion in Table 1., is used to develop the normalised decision matrix of TOPSIS (Table 2.) where, we have considered the range of alternatives for EV, ICEV and HEV in different price brackets.

| Alternatives | Cost of Ownership | Safety & Comfort | Range | Network Effect | Re-fuelling Infrastructure & Convenience | Environmental Impact | Policy Push & Regulations |
|---|---|---|---|---|---|---|---|
| **EV (8-11 Lakhs)** | 0.048 | 0.007 | 0.022 | 0.016 | 0.034 | 0.019 | 0.130 |
| **EV (11-15 Lakhs)** | 0.063 | 0.014 | 0.027 | 0.032 | 0.034 | 0.019 | 0.130 |
| **EV (15-19 Lakhs)** | 0.083 | 0.021 | 0.029 | 0.032 | 0.034 | 0.019 | 0.130 |
| **EV (19-25 Lakhs)** | 0.105 | 0.021 | 0.033 | 0.047 | 0.034 | 0.019 | 0.130 |
| **ICEV (8-11 Lakhs)** | 0.098 | 0.014 | 0.033 | 0.032 | 0.034 | 0.003 | 0.019 |
| **ICEV (11-15 Lakhs)** | 0.119 | 0.014 | 0.033 | 0.032 | 0.034 | 0.003 | 0.019 |
| ICEV (15-19 Lakhs) | 0.140 | 0.021 | 0.041 | 0.047 | 0.034 | 0.003 | 0.019 |
| ICEV (19-25 Lakhs) | 0.140 | 0.024 | 0.041 | 0.047 | 0.034 | 0.003 | 0.019 |
| HEV (19-25 Lakhs) | 0.114 | 0.021 | 0.049 | 0.016 | 0.034 | 0.005 | 0.093 |
| Vj+ | 0.048 | 0.024 | 0.049 | 0.047 | 0.034 | 0.019 | 0.130 |
| Vj- | 0.140 | 0.007 | 0.022 | 0.016 | 0.034 | 0.003 | 0.019 |

Table 2. Weighted Normalised Decision Matrix for BWM-TOPSIS

We obtain the positive ideal solution (PIS) and negative ideal solution (NIS) w.r.t. each criterion for different alternative in Table 2. The PIS and NIS are further used to calculate the separation measure for PIS and NIS, which is further used to compute the performance measure ($P_i$) for each of the available alternatives. The $P_i$ for each alternative is used to determine the rank of each alternative as the desired preference of the consumers.

| Alternatives | Si+ | Si- | Pi | Rank |
|---|---|---|---|---|
| EV (8-11 Lakhs) | 0.043 | 0.117 | 0.732 | 1 |
| EV (11-15 Lakhs) | 0.042 | 0.105 | 0.714 | 2 |
| EV (15-19 Lakhs) | 0.051 | 0.093 | 0.645 | 3 |
| EV (19-25 Lakhs) | 0.067 | 0.082 | 0.550 | 4 |
| HEV (19-25 Lakhs) | 0.091 | 0.049 | 0.382 | 5 |
| ICEV (8-11 Lakhs) | 0.104 | 0.032 | 0.349 | 6 |
| ICEV (19-25 Lakhs) | 0.117 | 0.045 | 0.284 | 7 |
| ICEV (15-19 Lakhs) | 0.117 | 0.046 | 0.276 | 8 |
| ICEV (11-15 Lakhs) | 0.086 | 0.053 | 0.238 | 9 |

Table 3. Separation Measure, Performance Score and Rank using BWM-TOPSIS

The results clearly indicate that EVs should be the preferable consumer choice over ICEVs when all the factors are considered for vehicle selection. According to our results in Table 3., it is clearly evident that EVs in the price bracket of 8.00-11.00 lakhs should be the first choice of the consumers. While ICEVs in the same price bracket is ranked as the sixth preferrable choice of consumers for vehicle purchase using BWM-TOPSIS . EVs in the price bracket 11.00-15.00 lakhs is ranked as the second desirable preference of the consumers while ICEVs of the same price segment are ranked as the ninth desirable preference of vehicle purchase for the buyers using BWM-TOPSIS . EVs in the price bracket 15.00-19.00 lakhs is ranked third desirable preference and its conventional counterpart is ranked as eighth desirable preference uing Table 3., . It has also been found that EVs in the price bracket of 19.00-25.00 lakhs should be the fourth desirable preference of the buyers from the available alternatives. Also, HEVs in the price bracket of 19.00-25.00 lakhs should be the fifth desirable preference of the buyers from the available alternatives while, ICEVs in the same price bracket are ranked as the seventh desirable preference of the buyers using Table 3. This indicates that HEVs should be preferred over ICEVs irrespective of the price segment. The findings that were obtained through the utilization of hybrid MCDM model using BWM and TOPSIS assist us in comprehending the fact that when EVs, ICEVs, and HEVs are to be selected from a price range that is comparable, EVs should be the first preferred choice of the consumers, while HEVs and ICEVs should be their second and third preferred choices, respectively.

Figure 3., shows us that how EVs, ICEVs and HEV from different price brackets is better than the other based on each factor considered in the study. The graph clearly indicates that TCO of EV is far less than the TCO of ICEVs and HEV in the same price band. Similarly, the policy push and regulations by the government are EV friendly as compared to ICEV and HEV. Also, the EVs are far more eco-friendly as compared to their conventional counterparts.

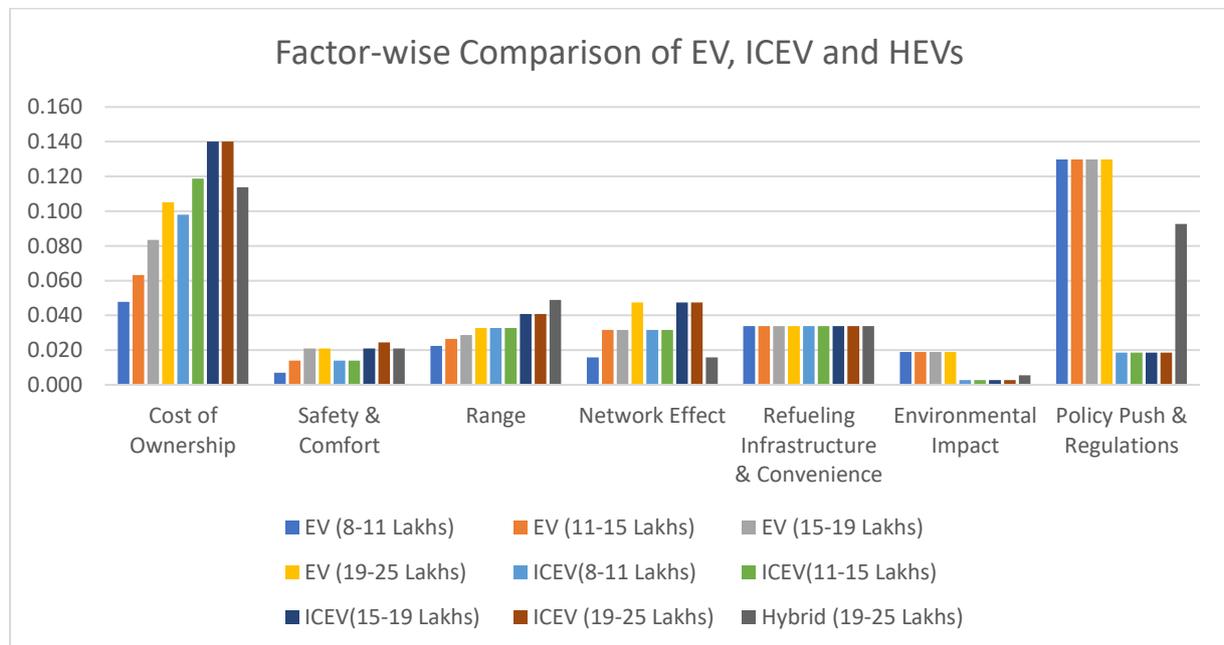

Figure 3.

## 9. Conclusion

EVs have enormous potential of becoming the mode of future mobility thus, enacting to save earth from various calamities that could occur due to global warming. They are a viable alternative for fossil fuel-based vehicles. This article discusses the impact of EVs over ICEVs in detail. The study has helped us to understand the fact that owning the electric vehicle is much cheaper than owning ICEV even though the initial purchase cost of EVs is much higher than that of ICEV. This is due to lower maintenance and operational cost of EVs throughout their life. The operating cost and initial purchase cost of hybrid vehicles is much higher than that of EVs. But the TCO for hybrid vehicles is lower than that of ICEVs of the same price bracket. If the factors in our study are considered for vehicle purchase, HEV must become the preferred choice of the buyers over ICEVs but not over EVs. From the study it is evident that EVs are more concerned towards the environment since they have zero tailpipe emission and also, they utilize the energy more efficiently as compared to traditional ICEVs. Also, the energy utilization by the hybrid vehicles is more judicious as compared to that of ICEVs but not judicious than EVs. The MCDM methodology used in our study i.e. BWM-TOPSIS have helped us infer the fact that EVs should be the first choice of the consumers followed by HEVs and ICEVs respectively.

Though EV adoption in India is rising at low pace majorly due to the lack of re-fuelling infrastructure and lower variety in the available EV segment. The government and the companies are taking steps to enhance the EV infrastructure and provide greater variety of EV models. Once the refuelling infrastructure is developed in the country and a greater number of EV models will come in the market (within the reach of consumers pocket), it will make a psychological impact on vehicle buyers to purchase EV rather than ICEV. Driving an EV provides safety and comfort due to the availability of regenerative braking system and contactless transmission. Thus, it can be concluded that EVs are better than their conventional counterparts (ICEVs) and hybrid vehicles so that, their adoption will enhance in near future. This will provide better quality automobiles with lower financial burden on pockets, enhanced safety to road accidents and also, enhance the climatic conditions in urban areas. Thus, EVs can be considered as the sustainable mode of transportation.